\documentclass[doublecol,showpacs,preprintnumbers,amsmath,cite]{epl2}
\usepackage[normalem]{ulem}
\usepackage{graphicx}
\usepackage{dcolumn}
\usepackage{bm}
\usepackage{amsmath}
\usepackage{amssymb}
\usepackage{color}

\begin{document}

\title{Shear Jamming in Granular Experiments without Basal Friction}

\author{Hu Zheng\inst{1,2,4}, Joshua A. Dijksman\inst{3,4} \and
R. P. Behringer\inst{4}}

\shortauthor{H. Zheng \etal}

\institute{
\inst{1} School of Earth Science and Engineering, Hohai University, Nanjing 210098, P. R. China\\
\inst{2} Department of Geotechnical Engineering, Tongji University, Shanghai 200092, P. R. China\\
\inst{3} Department of Physical Chemistry and Colloid Science, Wageningen University, Wageningen, The Netherlands\\
\inst{4} Department of Physics \& Center for Nonlinear and Complex Systems, Duke University, Durham, NC 27708, USA\\}

\abstract{Jammed states of frictional granular systems can be induced by shear strain at densities below the isostatic jamming density ($\phi_c$). It remains unclear, however, how much friction affects this so-called shear-jamming. Friction appears in two ways in this type of experiment: friction between particles, and friction between particles and the base on which they rest. Here, we study how particle-bottom friction, or basal friction, affects shear jamming in quasi-two dimensional experiments. In order to study this issue experimentally, we apply simple shear to a disordered packing of photoelastic disks. We can tune the basal friction of the particles by immersing the particles in a density matched liquid, thus removing the normal force, hence the friction, between the particles and base. We record the overall shear stress, and particle motion, and the photoelastic response of the particles. We compare the shear response of dry and immersed samples, which enables us to determine how basal friction affects shear jamming. Our findings indicate that changing the basal friction shifts the point of shear jamming, but it does not change the basic phenomenon of shear jamming.}

\date{\today}


\newcommand{\RNum}[1]{\uppercase\expandafter{\romannumeral #1\relax}}
\maketitle


Granular materials, which exhibit a great number of intriguing properties, have attracted much scientific attention in
recent years~\cite{geng04, hecke10, zhang10}. For example, granular
materials can turn from a loose fluid-like state into a
stress-supporting solid upon increasing the density of particles per
unit volume, a phenomenon called jamming~\cite{Liu1998,Hern2003}. Liu
\emph{et al}~\cite{Liu1998,Liu2010} proposed a jamming phase diagram
to capture the various state variables that determine whether a
material is jammed or not. The diagram was hypothesized to describe
not just the behavior of granular materials, but a whole range of
disordered materials, among which are colloids, foams and
emulsion. Granular materials in particular live on the
zero-temperature~$(T=0)$ plane of the Liu-Nagel jamming diagram, since
thermal fluctuations do not affect the macroscopic behavior of the
particles. Recently~\cite{bi11} it was shown that in this plane, the
phase behavior of granular materials is richer than the original phase
diagram suggested. Granular materials exhibit a property called shear
jamming, in which the simple shear deformation of a disordered
stress-free packing can turn it into a rigid structure, without
significantly changing the structure, an aspect not
covered by the work of Liu and Nagel.

This shear jamming phenomenon is apparent in studies of quasi two
dimensional (2D) photoelastic disk packings~\cite{bi11,ren13}. In
these systems, inter-particle forces were visualized with
photoelasticity~\cite{howell99,hartley03}, a technique that shows
clear force chain structures~\cite{majmudar05,majmudar07}. In these
systems friction appears in two ways: friction between particles, and
friction between particles and the base on which the particles rest,
i.e. \emph{basal} friction.Interparticle friction
  plays an important role in facilitating shear jamming.  But, in a
typical 2D photoelastic experiment one tries to reduce basal friction
by using powder-based lubricants. Still, it is impossible to totally
remove basal friction with lubricants. Usually, the basal friction is
assumed too small to substantially affect the results of the 2D
experiments \cite{majmudar05,majmudar07}. The ratio between fully
mobilized basal friction $F_f$ and mean contact force between
particles $F_p$ at shear jamming state is $\sim 0.12$, indicating that
basal friction should have little effect in determining the stresses
near or above jamming~\cite{bi11}. Floating particle systems have been
made before with airflow~\cite{puckett13} and with
water~\cite{JiePrivate}. However, we are not aware of experimental
studies probing directly the role of basal friction on shear jamming,
and choose a water based system for its experimental simplicity.

Here, we describe a novel apparatus that allows us to
eliminate static basal friction for the shear of quasi-2D photoelastic
disk packings, while effectively keeping all other
  experimental settings the same. We use this apparatus to
compare shear jamming for a basal-friction-free
particle packing, to a packing with basal
friction. We find that shear jamming persists in the absence of basal
friction. Eliminating the basal friction reveals two distinct
responses of the particle packing, which we associate with fragile and
shear jammed states~\cite{bi11}. We discuss the difference of their
responses via their change in shear stress, and their
different response visible in the deformation field of the
packing. We also compare the response of the basal friction
  free system to results obtained earlier in a shear setup with an
  articulated base, and find very similar phenomenology.

\section{Setup}
\label{sec:setup}

\begin{figure}[t!]
    \begin{center}
        \includegraphics[width=8cm]{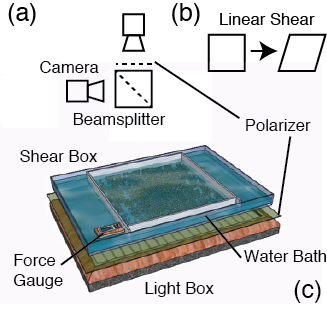}
        \caption{(a) Schematic picture of the camera setup, positioned
          above the bath with particles. The two-camera setup with
          non-polarizing beam splitter and circular polarizer for one
          camera, allows to record the particle displacements and
          photoelastic response simultaneouly. The distance between
          camera and bath is about 2 meters. (b) Isochoric simple
          shear deformation used in the experiments. (c) the water
          bath above the light box and polarizer. The force gauge is
          attached to the corner of the moving wall of the shear
          setup. \label{fig:setup}}
    \end{center}
\end{figure}

The apparatus consists of a 2 $\times~60$ $\times~80$~cm$^3$ shallow
water-tight tank, in which we have mounted an aluminum shear cell
(Fig.~\ref{fig:setup}a). One wall of the shear cell is driven by a
linear stage; bearings on the walls are constructed in such a way that
wall movement results in shear at constant volume (here, area) (40~cm
$\times 40$~cm) (Fig.~\ref{fig:setup}c). The walls all are rough on a
particle lengthscale.  We image the packing with two cameras via a
beam splitter to record both position and photelastic response of the
particles -- see Fig.~\ref{fig:setup}b. We rely on the system-averaged
squared intensity gradient $G^2 = |\nabla I|^2$ to serve as a proxy
for the stresses present in the system. Additionally, a force sensor
is positioned between the stage and the moving walls to record the
shear force, $F(\gamma)$, exerted on the particles during the imposed
deformation.  The stage is driven by a stepper motor at a speed of
approximately 0.33mm/s in all experiments unless otherwise stated. At
this speed, viscous stresses between the floating
particles are most likely negligible given the
  surface roughness of the particles and other imperfections in the
  line contacts the particles make. Rate-dependence of the
  dry-frictional interactions between them are also
weak~\cite{hartley03}.  The particles are photoelastic, custom made
from polyurethane sheets (Vishay PSM-4). In all our experiments, there
are about 3000 particles in the system. The particles are all
of uniform thickness ($\sim$ 6~mm), with three
different diameters:~$D_l=8.76 mm$, $D_m= 7.44 mm$ and $D_s=6.00
mm$. The number ratio of the large, medium and small (L:M:S) particles
is 5:22:4. The force sensor measurement necessarily measures both the
ensemble force of the packing and stray frictional forces (0.5-2~N)
from the sliders and bearings used to guide the motion of the walls;
the latter is very reproducible~\cite{Hu13} and is subtracted with a
calibration run. Particles float in a solution of ~$4 \%$ KCl in
demineralised water; the particles are just lighter than the salt
water and do not stick out of the liquid surface. They are thus not
affected by surface tension. The particles are also always
  millimeters separated from the base, making viscous forces between
  the particles and the base during particle motion much smaller than
  the experimental force resolution. In both wet and dry experiments,
  we start all experiments in a stress free state. The dry system is
  prepared stress free by gently ``massaging" the
    particle positions to remove as much photo-elastic signal as
  possible. This massaging will naturally induce frictional base
  loading in the arbitrary directions in which this massaging was
  necessary. This technique applied here is identical to the one used
  earlier~\cite{ren13}; we have never been able to detect any bias
  introduced by this method. Further details of the experimental
setup can be found elsewhere~\cite{Hu13}.

\section{Shear Without Basal Friction}\label{sec:sans}

\begin{figure}[t!]
    \begin{center}
        \includegraphics[width=8cm]{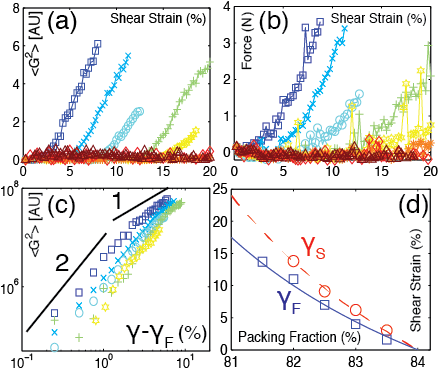}
        \caption{Wet system response. (a) $G^2(\gamma)$ for different $\phi$ as indicated
          with dark blue~$\square$: 83.5; light blue~$\times$: 83;
          cyan~o: 82.5; green~$+$: 82; yellow~$\ast$ 81.5;
          orange~$\divideontimes$: 81; red~$\lozenge$: 80.5; dark
          red~$\bigtriangleup$: 80\%. (b) Shear force $F(\gamma)$ for
          different densities; color, symbol legend identical to
          (a). (c) Double-logarithmic (base 10) plot of data from panel
          (a). Note the removal of the irrelevant large prefactor in (a). Indicated are the two regimes; solid lines indicate
          slopes 1 and 2 for reference. (d) Phase boundaries
          $\gamma_S, \gamma_F$ extracted from the $G^2$ data; the
          lines are fits. See text for details. \label{fig:fig2}}
    \end{center}
\end{figure}

To measure shear jamming in a basal friction free system, we apply
quasi-static isochoric shear to a floating layer of disks, which we
refer to as a \emph{wet} packing. We record the photoelastic signal
and shear force of the wet packing as a function of strain at a range
of packing fractions $\phi$. {\em Photoelastic pressure signal -- }
Due to residual stresses inside the particles, illumination
inhomogeneities and light refraction/scattering from particle edges,
any image will have a small offset $G_0^2$, even in the stress free
initial state. Since this background does not change during a given
run, we probe the photoelastic response by measuring $G(\gamma)^2 =
G_{raw}(\gamma)^2 - G_0^2$. $G^2(\gamma, \phi)$, as shown in
Fig.~\ref{fig:fig2}a. For all $\phi$ that the photoelastic response
emerges after a finite amount of strain $\gamma_F(\phi)$; the response
is initially super-linear $G^2(\gamma) \sim \gamma^{\beta}$ with
$\beta \gtrsim 1$; after some larger finite amount of strain,
$\gamma_S$, it evolves to linear behavior. The {\em shear force data}
measured from the force sensor shows the same trends as the
photoelastic response, as in Fig.~\ref{fig:fig2}b. To extract the
exponent $\beta$ and the strain amplitude $\gamma_F(\phi)$ signifying
the emergence of a force response, we plot $G^2$ vs. $\gamma$ on
double-logarithmic scales in Fig.~\ref{fig:fig2}c. We subtract from
each data set the $\gamma_F(\phi)$ which produces the best straight
line on a log-log plot. This method is very sensitive to
  small errors and hence is sufficiently accurate to extract a value
  for $\gamma_F$. For small strains beyond the noise plateau in the
$G^2$ data we observe $\beta \sim 1.8 \pm 0.3$.  We extract the
nonlinear to linear crossover point $\gamma_S(\phi)$ by extrapolating
the linear response regime to $G(\gamma_S) = 0$ on the
  linear scale from Fig.~\ref{fig:fig2}a, which yields results for
  $\gamma_S$ accurate to less than a percentage point strain. There
are two obvious limits for the functional behavior of $\gamma_F$ and
$\gamma_S$ with $\phi$: at lower $\phi$, no amount of shear can
shear-jam the packing. Below some threshold packing fraction $\phi_S$
we therefore expect $\gamma_{F,S} \rightarrow \infty$. At the
isotropic jamming point $\phi = \phi_J$, we expect $\gamma_{F,S} \rightarrow 0$. A simple function~\cite{expnote}
with these properties is:

\begin{equation}\label{eq:critgamma}
\gamma_{F,S} = \gamma^C_{F,S}\frac{\phi_J - \phi}{\phi - \phi_S}
\end{equation}

We plot $\gamma_F(\phi)$ and $\gamma_S(\phi)$ in
Fig.~\ref{fig:fig2}~d, and show that they delineate two phase
boundaries which merge at the isotropic jamming point $\phi_S \sim
84\%$.  We then fit the data for $\gamma_{F,S}$, drawn from the $G^2$
data, to Eq.~\ref{eq:critgamma}, where we use $\phi_S = 75\%$, based
on the present experiments, and $\phi_J = 84\%$ from Ref.~\cite{bi11}.
The amplitudes$\gamma^C_{F,S}$ are different fit parameters for the
two cases. The fits are shown as lines in Fig.~\ref{fig:fig2}d, and
are good representations of the data. We find that $\gamma^C_F \simeq
35\%$ and $\gamma^C_S \simeq 50\%$. We also extract $\gamma_S$ from
the shear force data~\cite{footnote2} by extrapolating the linear
response regime to $F(\gamma_S) = 0$. The results for $\gamma_S$ (not
shown) is consistent with  corresponding data obtained from $G^2$.

\section{Comparison to Shear with Basal Friction} 
\label{sec:with}

\begin{figure}[t!]
    \begin{center}
        \includegraphics[width=8cm]{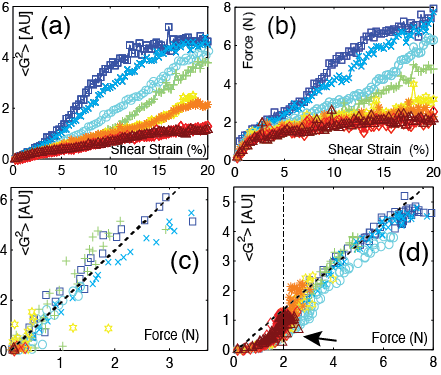}
        \caption{Dry system response (a,b). (a) $G^2(\gamma)$ for different $\phi$ as indicated
          with different colors and symbols: dark blue~$\square$:
          $\phi = 83.5$; light blue~$\times$: 83; cyan~o: 82.5;
          green~$+$: 82; yellow~$\ast$ 81.5; orange~$\divideontimes$:
          81; red~$\lozenge$: 80.5; dark red~$\bigtriangleup$: 80\%
          . (b) Shear force $F(\gamma)$ for different densities;
          color, symbol legend identical to (a). (c) Direct comparison
          of the wet packing data in Fig.~\ref{fig:fig2}a,b via $G^2(F)$. The
          dashed line indicates a linear correlation. (d) $G^2(F)$ for
          dry packing experiments shown in panel a,b. The dashed line
          is a best linear correlation. The arrow indicates an excess
          of shear force measured that does not reveal itself in the
          photoelastic response. The dash-dotted line indicates the
          maximum shear force for fully mobilized frictional contacts
          (see text).
        \label{fig:fig3}}
    \end{center}
\end{figure}   

Above, we have characterized shear jamming and relaxation in a wet
packing. We next determine the role of basal friction. By not having
liquid in our setup, we ``turn on'' friction with the base, while
keeping all other experimental settings the same. We perform the same
shear experiments as in the wet case, but with particles resting
directly on the acrylic base plate of the tank, which we call the
\emph{dry} packing. We normalize the photoelastic response data in the
same way as the wet system~\cite{footnote1}.\\

We summarize the response under dry conditions in
Fig.~\ref{fig:fig3}a,b. In part a, we show the photoelastic response,
which first increases linearly, even for the lowest packing fraction
considered here ($\phi = 80.0\%$).  For higher densities, we observe a
sharp increase in the slope, $d G^2/d \gamma$, similar to the
basal-friction-free system, followed by saturation in
$G^2(\gamma)$. In Fig.~\ref{fig:fig3}b, we show the force sensor data
obtained during the same runs. For large and intermediate strains we
observe similar trends to the photoelastic response (although the
force saturation does not happen for $\phi < 83.5\%$). For lower
$\phi$, we see that the force sensor data deviates from the
photoelastic response: at $\phi = 80.0\%$, the shear force increases
substantially for small strains, and then saturates.

Comparing Fig.~\ref{fig:fig2}a,b with   Fig.~\ref{fig:fig3}a,b, there are three characteristics in the dry   packing dynamics that are absent from the wet packing dynamics: \textit{(i)} $G^2$ increases slowly with strain, even for the lowest
packing fraction $\phi = 80\%$; \textit{(ii)} for $80\% \leq \phi
  \leq 80.5\%$, $F$ increases relatively quickly, but for these low
$\phi$'s, $F$ saturates at $\gamma \sim 3\%$ to about $F \sim 2N$;
\textit{(iii)} $G^2, F$ for our largest $\phi = 83.5\%$ exhibit a plateau at large strain.

We gain insight into the physical origin of observations \textbf{i -- iii} by looking first at the correlation between $G^2(F)$ for both the wet and dry system -- Fig.~\ref{fig:fig3}c, d respectively. Fig.~\ref{fig:fig3}c shows that the photoelastic response and the shear force in the wet packing are linearly correlated. However, for the dry system, this linear correlation is not as good, as shown in Fig.~\ref{fig:fig3}d. At small $F$, where also the applied strains are small, the photoelastic response in the dry packing increases much less than expected based on a best linear fit (dashed line). We attribute this excess of shear force to increased mobilization of frictional contacts with the base, with the following reasoning. For the dry case, the applied force, $F$, is balanced by three other types of forces. One of these is friction in the apparatus, which is subtracted. The second is due to inter-particle contact forces which are ultimately balanced by forces at the boundaries. The third is due to friction between the particles and the base. Before strain is applied, basal friction forces are mobilized in arbitrary directions. With each successive strain step, basal friction forces, which have minimal effect on the photoelastic response, become mobilized so as to resist the applied force, which is applied through interparticle contacts, hence the roughly linear increase in $G^2$, explaining observation \textit{i}. This mobilization effect saturates at a maximum force of $\sim 2N$, because maximum mobilization of the basal contact forces is reached; beyond this point static friction fails, contact forces cannot grow anymore, and particles start to move.

To support this view, we quantify the basal friction effect, by
measuring the maximum friction to move the whole system on the
Plexiglas $f_{max}=\mu m g = \mu\rho\phi A H g$. Here, $\mu$ is the
coefficient of friction between particle and Plexiglas, g is the
gravitational acceleration, $H$ is the disk height and $A_i$ is the
systems's cross sectional area. The maximum force, when friction is
fully mobilized, for $\phi=80\%$ is $f_{max} \simeq 2~N$, which yields
a $\mu = 0.24$, which is low but not unreasonable for the
powder-lubricated particles resting on the plexiglas
base --- it appears that for $\phi \le 80.5\%$, the system
  does not actually jam. This phenomenon explains
  observation\textit{(ii)}. For observation \textit{(iii)}, the
  situation is more complicated.  One possible explanation of the
  saturation of $G^2$ is that the photoelastic response saturates at
  large pressure. However, this does not account for the essentially
  identical saturation in $F$. In this regard, we note two possible
  sources of saturation: a) shear bands develop, which limit the
  maximum attainable shear stress; and b) even when shear bands do not
  form, as in the experiments of Ren et al.\cite{ren13}, for large
  enough strains, well above shear jamming, the system simply becomes
  for isotropic. We are currently carrying out additional shear
  experiments on particles with much lower friction coefficients to
  test these ideas. Note also that the proportionality between $G^2$
  and $F$ is different for the dry and wet case. This calibration
  factor depends on several experimental details, such as light
  intensity and camera aperture, which may have been different in the
  two experiments.

\subsection{Particle Tracking}

\begin{figure}[t!]
    \begin{center}
        \includegraphics[width=8cm]{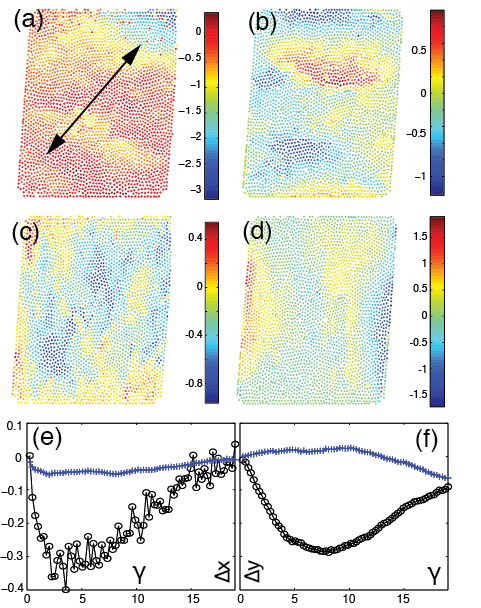}
        \caption{(a) Non-affine deviation from linear shear in the
          horizontal displacement of individual particles for a dry,
          82\% packing sheared to $\gamma = 7.6\%$. Color scale
          indicates displacement magnitude in average particle radius;
          the arrow indicates the dilation direction. The outliers visible in panel (a) are tracking artifacts. (b) As (a), but
          for a wet packing at the same packing fraction. 
          (c,d) as for (a,b), respectively, but now showing the vertical non-affine displacements of the particles.
          (e) The mean horizontal non-affine displacement per particle. (f) Mean
          vertical non-affine displacement. For (e) and (f): Dry packing: black $o$, wet packing: blue $+$. Units are in mean
          particle radius for all panels.\label{fig:figtrack}}
    \end{center}
\end{figure}

We obtain additional evidence for the physical picture put forward
above, by probing the packing deformation and the displacement of
individual particles. In Fig.~\ref{fig:figtrack} we show the
non-affine~\cite{footnote3} horizontal particle displacements for both
a dry (a) and a wet (b) packing at 82 \% and strain amplitude of
$\gamma = 7.6\%$. In the dry packing, the top right part of the
packing is lagging the lower left: particles in the packing remain
static until the interparticle force overcomes basal
friction. A particle moves when it experiences a force
  which can overcome the resistance from its neighbor particles and
  the basal friction which it is experiencing; this leads to a local
  compaction of the packing. The horizontal non-affine
  response is indeed much more homogeneous, although
  large spatially coherent inhomogeneities can be observed. The part
of the box that is mobilized last is the corner in the expansion
direction of the shear, furthest away from any pushing wall. The
system-averaged mean non-affine horizontal displacements for the wet
and dry system, shown in Fig.~\ref{fig:figtrack}e
  supports this view: it is evident in the dry case that the
  non-affine particle displacements have an extreme value at a strain
  amplitude of $\gamma \simeq 3\%$. A similar lag, but with an
  extremum at $\sim 6\%$ occurs in the system-averaged mean non-affine
  vertical displacement, Fig.~\ref{fig:figtrack}f. The peak lag moment
  in Fig.~\ref{fig:figtrack}e corresponds to the point where the shear
  force saturates in Fig.~\ref{fig:fig3}b, where interparticle forces
  have grown to overcome basal friction. Once all basal contacts are
  mobilized, the shear force saturates because most of the packing is
  moving in a basically unjammed state for the lowest $\phi$'s, and
  the non-affine motion becomes random. In contrast, in the wet
packing, the non-affine displacements are almost equally positive and
negative, i.e. random, and the system-averaged mean non-affine motion
is only a fraction of that of the dry system. The source of
  this small amount of non-affine motion is elucidated by looking at
  the vertical particle displacement field, shown in
  Fig.~\ref{fig:figtrack}c,d for both the wet and dry system in the
  same frame as panel (a,b). For the wet system, they indicate a small
  amount of mass transport at the boundaries; the small inbalance in
  non-affine transport is thus a boundary effect, where perhaps
  density inhomogeneities drive this flow.

\section{Reynolds Pressure and Relaxation Effects}

\begin{figure}[t!]
    \begin{center}
        \includegraphics[width=8cm]{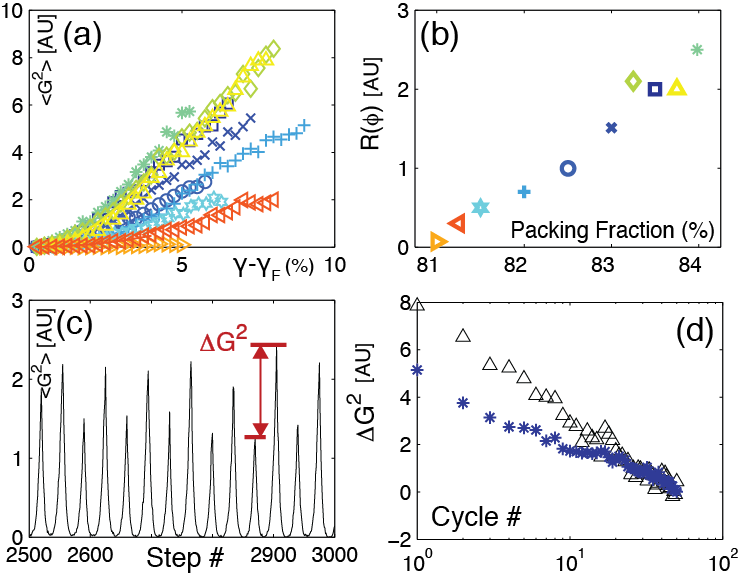}
        \caption{(a) Quadratic pressure increase after $\gamma_F$ at different packing fractions: dark green $\star$: $\phi = 83.98$; yellow 								$\bigtriangleup$: 83.75; dark blue~$\square$: 83.5; light green $\lozenge$: 83.25; light blue~$\times$: 83; grey~o: 82.5;
          cyan~$+$: 82; menthe~$\ast$ 81.75; dark orange~$\triangleleft$:
          81.5; light orange~$\triangleright$: 80.25. (b) For the first 3\% shear, a quadratic fit is also plotted for each run. (b) $R(\phi)$ for the fits shown in (a); symbols and colors are identical. (c) $\Delta G^2(n)$ for a $\phi = 83.75\%$ run with $n=70$ for each cycle. The window shown is after about 35 complete cycles; the response in $G^2$ is almost symmetric. Indicated is also the definition of $\Delta G^2$. (d) $\Delta G^2(n)$ for $\phi = 83.75\%$ (triangles) and $\phi = 83.89\%$ (stars). Note the logarithmic abscissa.       
        \label{fig:figJiecomp}}
    \end{center}
\end{figure}

In previous work~\cite{ren13}, we quantified Reynolds dilatancy and relaxation effects in a 2D packing driven by a Galilean invariant deformation through an articulated base. However, the interaction between the base and the particles was still frictional. Here we test if the observations from~\cite{ren13} reproduce also in the absence of basal friction. Although the experimental conditions of the two experiments are substantially different, we expect two key effects to be qualitatively similar: \textit{(i)} The prefactor for the quadratic rise in the pressure with strain should increase with the packing fraction. \textit{(ii)} The force response of the packing to asymmetric shear should logarithmically decay to a symmetric response. In this section, we demonstrate that these two key observations are indeed also observed in the absence of basal friction. In Fig.~\ref{fig:figJiecomp}a, we show the quadratic increase of the pressure after $\gamma_F$ for different packing fractions. Indeed, the prefactor increases with the packing fraction, as shown by Fig.~\ref{fig:figJiecomp}b, where we show the prefactor $R(\phi)$ from $G^2 = R(\phi)\gamma^2$ that we have called the Reynolds coefficient~\cite{ren13}. The divergence of the coefficient towards $\phi_J$ is not as strong as observed in previous work, but it does rise strongly with $\phi$.\\
Next, we compare the response of the wet system to cyclic shear. Before addressing this, we briefly review some details of the
protocol and response. One cycle of the protocol involves shear strain in the 'forward' direction in small steps to a maximum strain,
$\gamma_{max}$, followed by a reversal of the shear to the original system configuration, i.e. to $\gamma_{min} = 0$. The pressure, measured in $G^2$ reached a maximum at $\gamma_{max}$ in the first cycle. Associated with the growth of the pressure
is the evolution of an anisotropic strong force network, where the compressive direction of shear strain corresponded to the principal
direction of network. During reversal of the strain, we have seen before that $G^2$ and the original force network diminish, and a new strong network formed perpendicular to the first network, and associated with the fact that the compressive and dilational directions switch under shear strain
reversal. When the system returned to the unstrained configuration, a strong force network associated with the reverse
strain persisted at $\gamma = 0$. Associated with the network is a non-zero pressure. In subsequent cycles of forward and reverse shear,
the pressure at $\gamma_{max}$ decreased monotonically with cycle number, and the pressure at $\gamma = 0$ grew correspondingly. The
difference $G^2(\gamma_{max}) - G^2(\gamma_{min}) = \Delta G^2$ decreased logarithmically with cycle number $n$, suggesting an activated process in the stress ensemble, enabled through fluctuations in the interparticle forces during shear.  These observations reproduce very well in the basal friction free setup. In Fig.~\ref{fig:figJiecomp}c we show $G^2(n)$ for an experiment at $\phi = 83.75\%$ and $\gamma_{max} = 7.6\%$ after about 35 cycles. Clearly the force response has become almost symmetrical in this experiment. The decay of the asymmetry is also logarithmic, as in the articulated base experiment. We measure the asymmetry through $\Delta G^2(n)$ indicated in Fig.~\ref{fig:figJiecomp}c. The decay of $\Delta G^2(n)$ is clearly logarithmic as indicated in Fig.~\ref{fig:figJiecomp}d. We can thus conclude that even more subtle features of shear jamming, such as Reynolds dilatancy and pressure relaxation, are robust in the absence of basal friction~\cite{footnote4}.

\section{Conclusions}
We have studied the role of basal friction on the stress and flow dynamics of a sheared two dimensional packings of frictional
disks. Our main finding from the photoelastic response of the packing is that shear jamming and many associated subtle features also occur in the absence of basal friction. This observation is supported by two independent stress measurements and a comparison of the packing force response with that from an entirely different driving mechanism. We identify the onset of fragile and shear jammed states. Our findings have interesting repercussions: We demonstrate that a wall-driven floating particle system is superior to a similar system where the particles rest on a static frictional base. In particular, a floating system removes any effect due to gravity, making it ideal for micro-gravity granular studies.  The emergence of rigidity in our slowly sheared packing also hints that a simple frictional mechanism can be the sole source of the viscosity divergence of dense athermal suspensions~\cite{boyer11, seto13}. Interestingly, the flow fields observed in the dry and wet experiments are very different, despite their force response being very similar. This suggests a surprising disconnect between the applied stresses and resulting strains, which is of interest for constitutive modeling attempts. The analysis of these flow fields will be the subject of future work.

\section{Acknowledgements}
We thank Richard Nappi for help in the machine shop, Abe Clark, Junyao
Tang and Jie Ren for help with image analysis, and Kim Clark for
baking cookies. IGUS generously provided a linear stage under the
Young Engineers Support Program. This project was funded by NASA grant
NXX10AU01G, NSF grants NSF0835742, NSF-DMR12-06351, and ARO grant
W911NF-1-11-0110.  HZ was supported by the Chinese Scholarship Council
and a travel grant from Tongji University.

\bibliographystyle{prsty}

\begin{thebibliography}{10}
\bibitem{expnote}
The exponent to which the $\phi$-dependent numerator and denominator are raised, in Eq.~\ref{eq:critgamma}, can be of any order; one is the simplest that works. The equation is only assumed to be valid between $\phi_J$ and $\phi_S$.
\bibitem{footnote2} The static friction correction for the shear force is not accurate enough for the shear force data from the force gauge to aid in determining $\gamma_S, \gamma_F$.
\bibitem{footnote1}
To normalize the photoelastic response measured at $\phi = 83.5\%$, we use the $G_0^2$ measured at $83\%$, because at $83.5\%$, the initial state cannot be prepared stress free. This issue does not arise in the floating particle system.
\bibitem{footnote3} By non-affine we refer here to the displacements
  that remain after subtracting the mean shear strain, that is, the
  linear shear profile, interpolated to a given particle position.
\end{thebibliography}

\end{document}